\documentclass[a4paper,11pt]{article}
\usepackage{pos}
\usepackage{amsmath}
\usepackage{multirow}
\usepackage{array}
\usepackage{url}
\usepackage{graphicx}
\usepackage{microtype}
\usepackage{multicol}
\newcommand{\as}{\alpha_s}
\newcommand{\mz}{m_Z}

\newcommand{\gev}{\,\mathrm{GeV}}
\newcommand{\tev}{\,\mathrm{TeV}}
\newcommand{\ab}{\,\mathrm{ab}^{-1}}
\newcommand{\fb}{\,\mathrm{fb}^{-1}}

\title{QCD studies and precision physics at the LHeC}
\ShortTitle{Precision studies at the LHeC}

\author*[a,b]{Francesco Giuli}
\affiliation[a]{Universit\`a degli Studi Link,  Via del Casale di S. Pio V, 44, 00165\
Rome, Italy}

\affiliation[b]{INFN Sezione di Roma Tor Vergata, Via della Ricerca Scientifica, 1, 00133,\
Rome, Italy}

\abstract{%
The Large Hadron electron Collider (LHeC) would add a high-current Energy Recovery Linac to the HL-LHC, delivering electron--proton collisions at centre-of-mass energies around the TeV scale. This contribution summarises the QCD and parton-distribution-function (PDF) aspects of the recent LHeC bridge-project study. The combination of high luminosity, a very large lever arm in Bjorken $x$ and $Q^2$, and clean neutral- and charged-current deep-inelastic scattering measurements would enable a coherent determination of all proton PDFs in a single experiment. The resulting constraints would substantially reduce uncertainties in the gluon, valence, strange and heavy-flavour distributions, provide stringent tests of perturbative and small-$x$ QCD, and improve the parton luminosities that enter precision and discovery measurements at the HL-LHC and at future hadron colliders. The same programme gives competitive and complementary determinations of the strong coupling and weak mixing angle, including measurements of its running over a wide range of scales.}

\FullConference{The 33rd International Workshop on Deep Inelastic Scattering and Related Subjects (DIS2026)\\
4--8 May 2026\\
Bologna, Italy\\}

\begin{document}
\maketitle

\section{Introduction}

The LHeC is proposed as a high-energy deep-inelastic scattering (DIS) facility based on a $50\gev$ electron Energy Recovery Linac colliding with the HL-LHC proton and ion beams. In a dedicated post-HL-LHC running scenario, it can collect about $1\ab$ in $ep$ mode over six years, with a first-year dataset of order $50\fb$. The resulting centre-of-mass energy, $\sqrt{s}\simeq 1.2\tev$ for $ep$ collisions, extends DIS far beyond the HERA kinematic domain while retaining the clean experimental and theoretical environment of lepton--hadron scattering~\cite{Ahmadova:2025,Agostini:2020,LHeCStudyGroup:2012zhm,Klein:2018rhq}.

For QCD, this programme is distinctive in two related ways. First, neutral-current (NC) and charged-current (CC) inclusive DIS determine the flavour decomposition of the proton over a broad kinematic range with common detector systematics. Second, the same dataset constrains the proton structure needed for precision measurements and searches in hadron collisions. The LHeC therefore plays both a standalone role, by testing perturbative and small-$x$ QCD, and an enabling role, by improving the predictions used in the HL-LHC and future collider physics programmes.

\begin{figure}[t]
\centering
\includegraphics[width=0.72\textwidth]{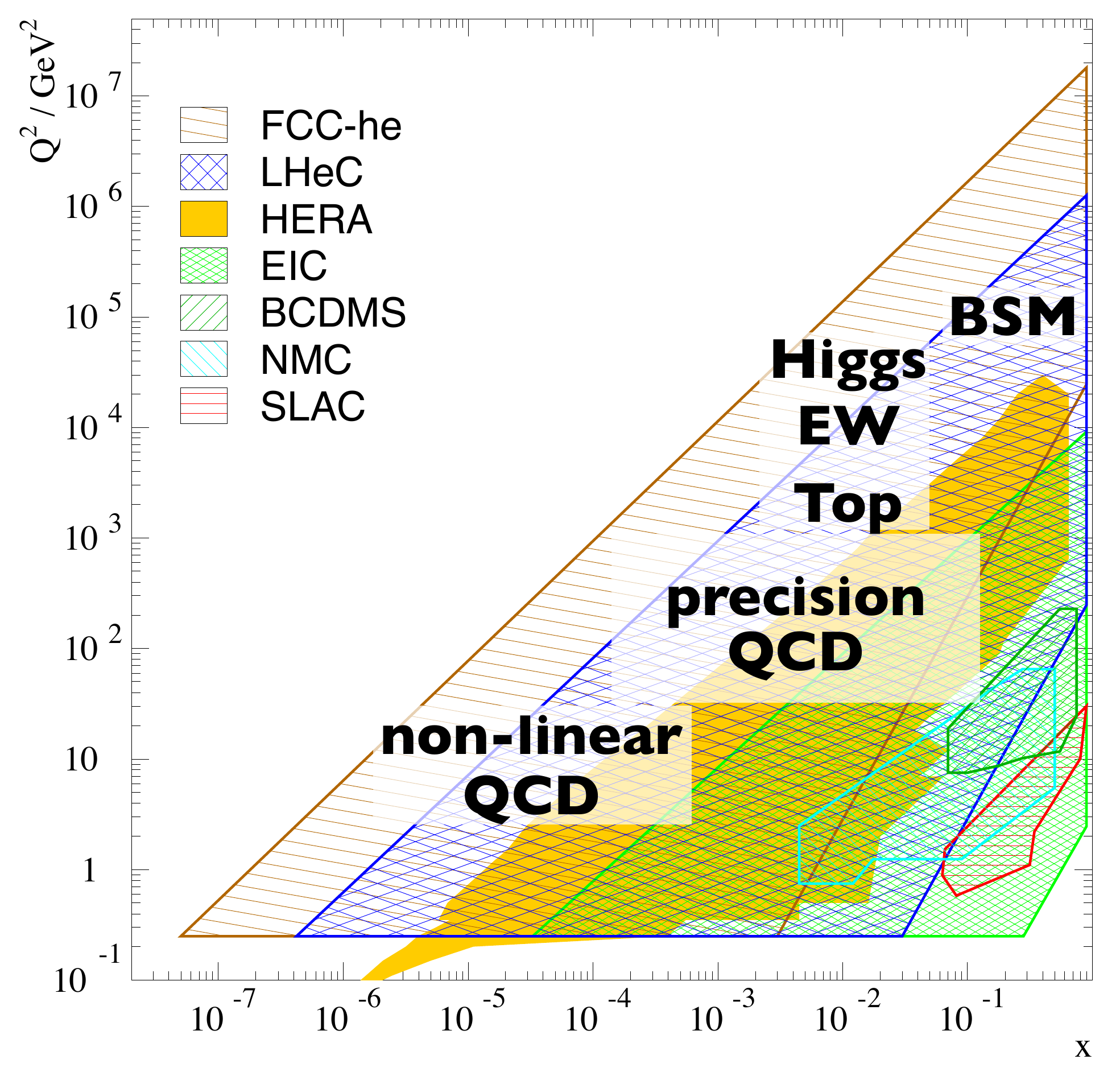}
\caption{Kinematic coverage in $(x,Q^2)$ for DIS experiments. The LHeC extends the HERA reach by several orders of magnitude in both the low-$x$ and high-$Q^2$ directions, entering regions relevant for precision QCD, electroweak, Higgs, top and BSM measurements.}
\label{fig:kinplane}
\end{figure}

\section{PDF constraints from LHeC DIS}

The central PDF observable is the set of NC and CC double-differential DIS cross sections. NC data constrain the quark singlet and gluon through scaling violations, while CC data provide direct flavour sensitivity through $W$ exchange. The motivation is aligned with the present global-PDF goal of reaching percent-level proton-structure accuracy for LHC phenomenology~\cite{NNPDF:2021njg,PDF4LHCWorkingGroup:2022cjn}. Compared with HERA, the larger centre-of-mass energy increases the accessible range in $x$ and $Q^2$ by several orders of magnitude relative to the combined HERA inclusive programme~\cite{H1:2015ubc}. This is important at both ends of the kinematic plane. At small $x$, the LHeC probes gluon densities in a region where non-linear dynamics, small-$x$ resummation, or other departures from fixed-order DGLAP evolution may become visible. At high $x$ and high $Q^2$, it gives direct DIS information in the region that controls high-mass tails at hadron colliders.

A key feature of the LHeC projections is that all light-quark combinations and the gluon can be determined in one experiment. This reduces the dependence on combining heterogeneous datasets with different normalisations, heavy-target corrections and theoretical assumptions. The projected PDF fits use NC and CC inclusive pseudodata, with semi-inclusive strange-sensitive information included for the $s+\bar s$ determination, and are compared to the current CT, MSHT, NNPDF and ABMP global analyses~\cite{Hou:2019efy,Bailey:2020ooq,Alekhin:2017kpj,NNPDF:2021njg}. Heavy-flavour final states extend the programme to charm and bottom and give sensitivity to possible intrinsic heavy flavour at large $x$~\cite{Agostini:2020,Ahmadova:2025}.

\begin{figure}[t]
\centering
\begin{minipage}{0.48\textwidth}
\centering
\includegraphics[width=\linewidth]{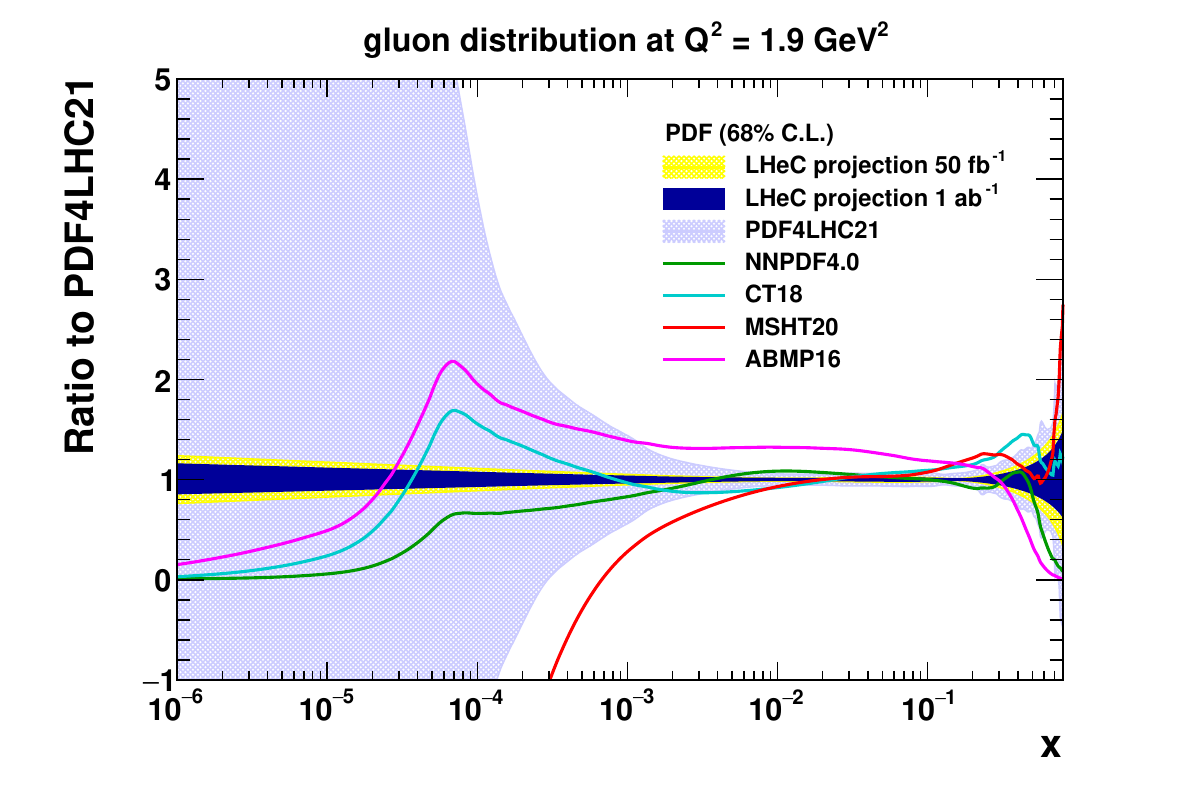}
\end{minipage}\hfill
\begin{minipage}{0.48\textwidth}
\centering
\includegraphics[width=\linewidth]{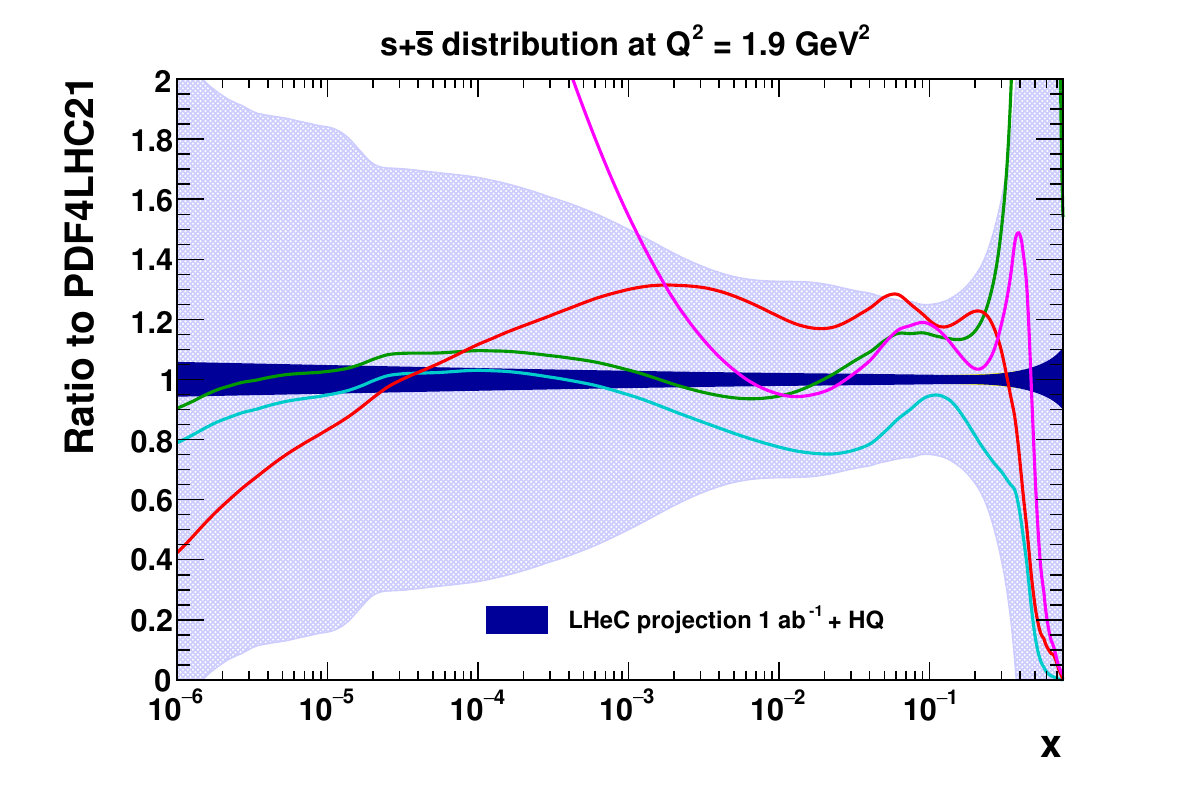}
\end{minipage}
\begin{minipage}{0.48\textwidth}
\centering
\includegraphics[width=\linewidth]{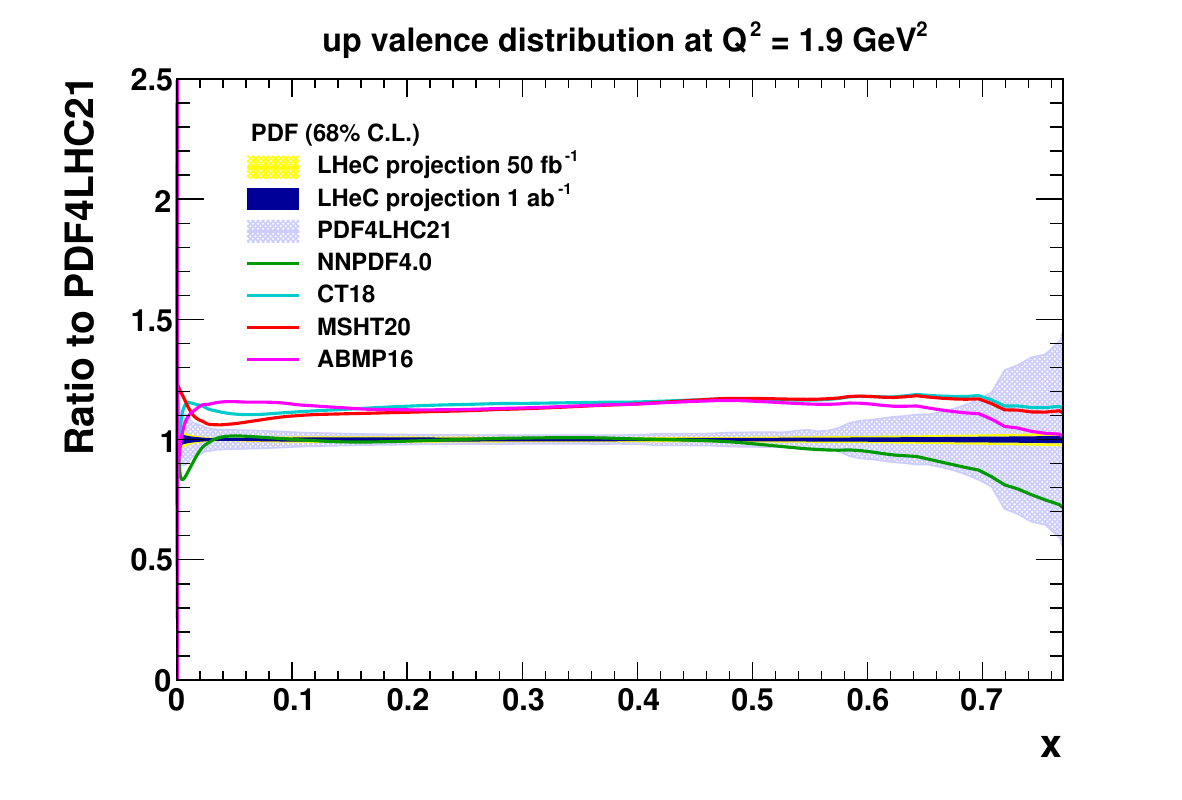}
\end{minipage}\hfill
\begin{minipage}{0.48\textwidth}
\centering
\includegraphics[width=\linewidth]{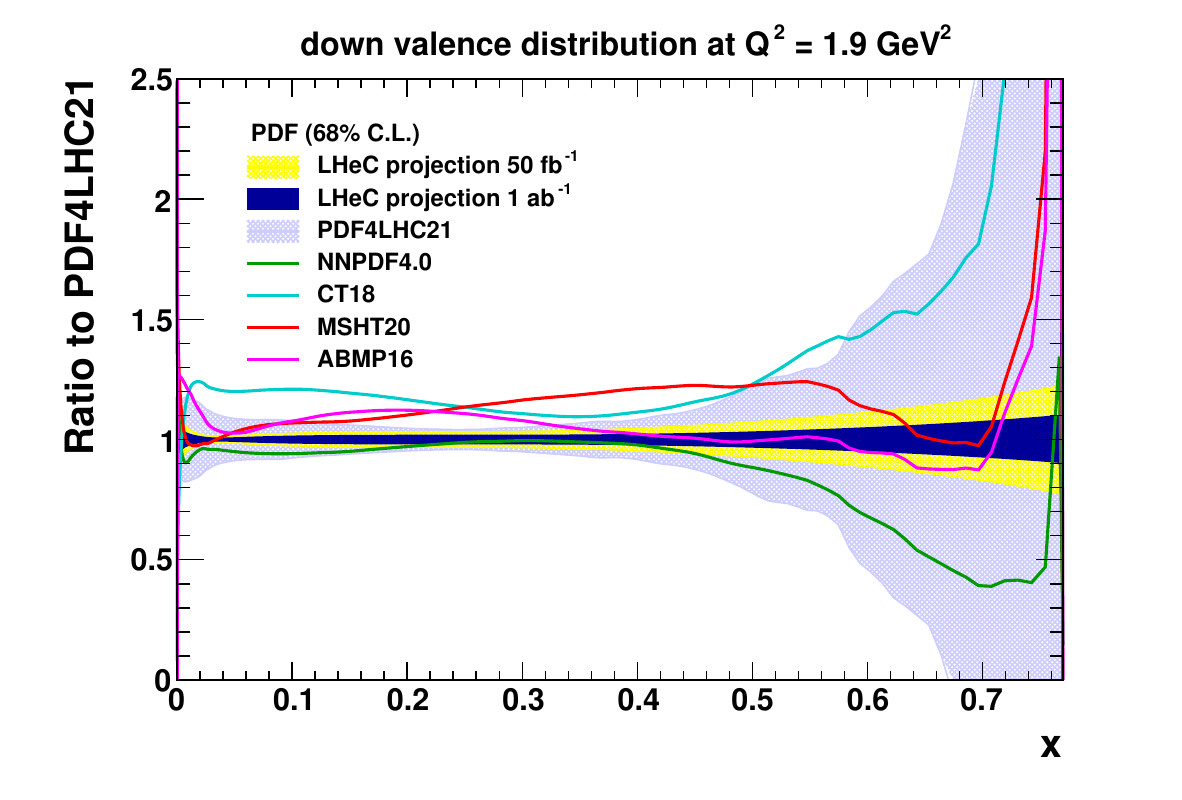}
\end{minipage}
\caption{Projected LHeC impact on representative PDFs at the starting scale $Q^2=1.9\gev^2$, shown as ratios to PDF4LHC21. The panels show, clockwise from top left, the gluon, $s+\bar s$, down-valence and up-valence distributions. The projected uncertainties for $50\fb$ and $1\ab$ are compared with the spread of current global analyses.}
\label{fig:pdfs}
\end{figure}

Fig.~\ref{fig:pdfs} illustrates the scale of the improvement. For the gluon, the LHeC constrains both the poorly known small-$x$ region and the high-$x$ tail, where present global fits differ substantially. The valence distributions become precise over the region most relevant to high-mass Drell--Yan and new-resonance searches. The strange sea is determined with much reduced uncertainty, providing an important input for electroweak precision analyses. The comparison of the $50\fb$ and $1\ab$ projections shows that even an early LHeC dataset would already have a sizeable impact, while the full dataset would make the PDF uncertainty bands much narrower than the present PDF4LHC21 envelope.

\section{Impact on hadron-collider parton luminosities}

The PDF constraints from LHeC data propagate directly to parton--parton luminosities at hadron colliders. This is crucial because precision Higgs, electroweak and top-quark measurements at the HL-LHC will increasingly be limited by proton-structure uncertainties, while searches at the largest invariant masses depend on extrapolations into regions where current PDFs are weakly constrained. The recent study finds that LHeC-improved uncertainties in relevant luminosities can be at least a factor of five smaller than those of PDF4LHC21, while also discriminating among the current global fits that enter the PDF4LHC combination~\cite{Ahmadova:2025,PDF4LHCWorkingGroup:2022cjn}.

\begin{figure}[t]
\centering
\begin{minipage}{0.48\textwidth}
\centering
\includegraphics[width=\linewidth]{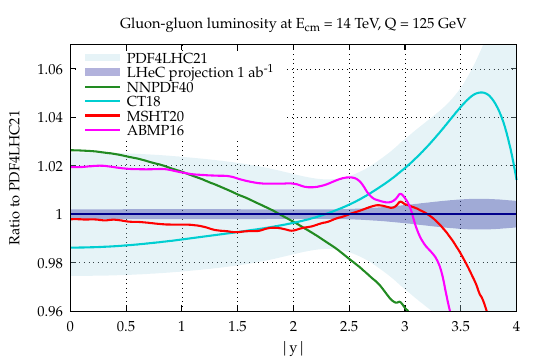}
\end{minipage}\hfill
\begin{minipage}{0.48\textwidth}
\centering
\includegraphics[width=\linewidth]{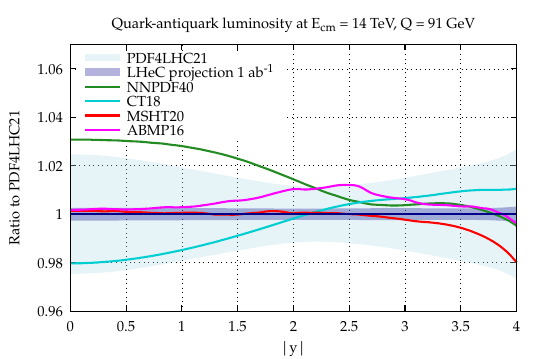}
\end{minipage}
\begin{minipage}{0.48\textwidth}
\centering
\includegraphics[width=\linewidth]{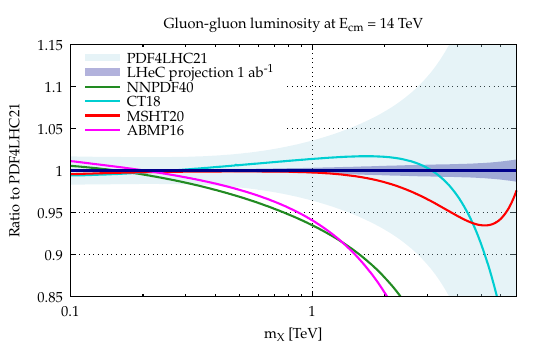}
\end{minipage}\hfill
\begin{minipage}{0.48\textwidth}
\centering
\includegraphics[width=\linewidth]{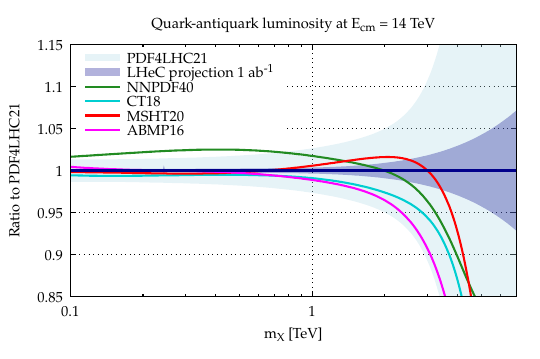}
\end{minipage}
\caption{Projected parton-luminosity impact at $\sqrt{s}=14\tev$, normalised to PDF4LHC21. Top: rapidity dependence of $gg$ luminosities at $Q=125\gev$ and $q\bar q$ luminosities at $Q=91\gev$. Bottom: invariant-mass dependence of the corresponding luminosities. The LHeC bands are much narrower than current global-PDF differences.}
\label{fig:lumi14}
\end{figure}

Fig.~\ref{fig:lumi14} gives representative examples for $\sqrt{s}=14\tev$. The rapidity dependence of $gg$ and $q\bar q$ luminosities is directly relevant for Higgs and Drell--Yan measurements, respectively. The invariant-mass dependence controls the normalisation and shape of high-mass spectra. Reducing the PDF component in these observables is a prerequisite for using the full HL-LHC statistics and for interpreting small deviations in tails, where BSM effects can interfere with Standard Model amplitudes and may otherwise be partially absorbed into fitted PDFs.

The same logic extends to future hadron colliders. At $100\tev$, the physics programme relies on PDFs at lower $x$ for Standard Model processes and at high $x$ for the highest-mass searches. A prior LHeC dataset would therefore be a key input to the FCC-hh or any future energy-frontier hadron collider, reducing both central-value ambiguities and PDF uncertainties before such a machine starts operation.

\section{Strong coupling}

The LHeC can determine $\as$ through two largely complementary methods. Inclusive NC and CC DIS constrain $\as$ through QCD scaling violations and its correlation with the gluon PDF. Jet production, especially in the Breit frame, gives a more direct hard-QCD observable. The detector concept, with high-granularity calorimetry and in-situ calibration from NC DIS, targets a jet-energy-scale uncertainty around the sub-percent level, which is essential for a competitive jet-based $\as$ extraction.

\begin{figure}[t]
\centering
\includegraphics[width=0.76\textwidth]{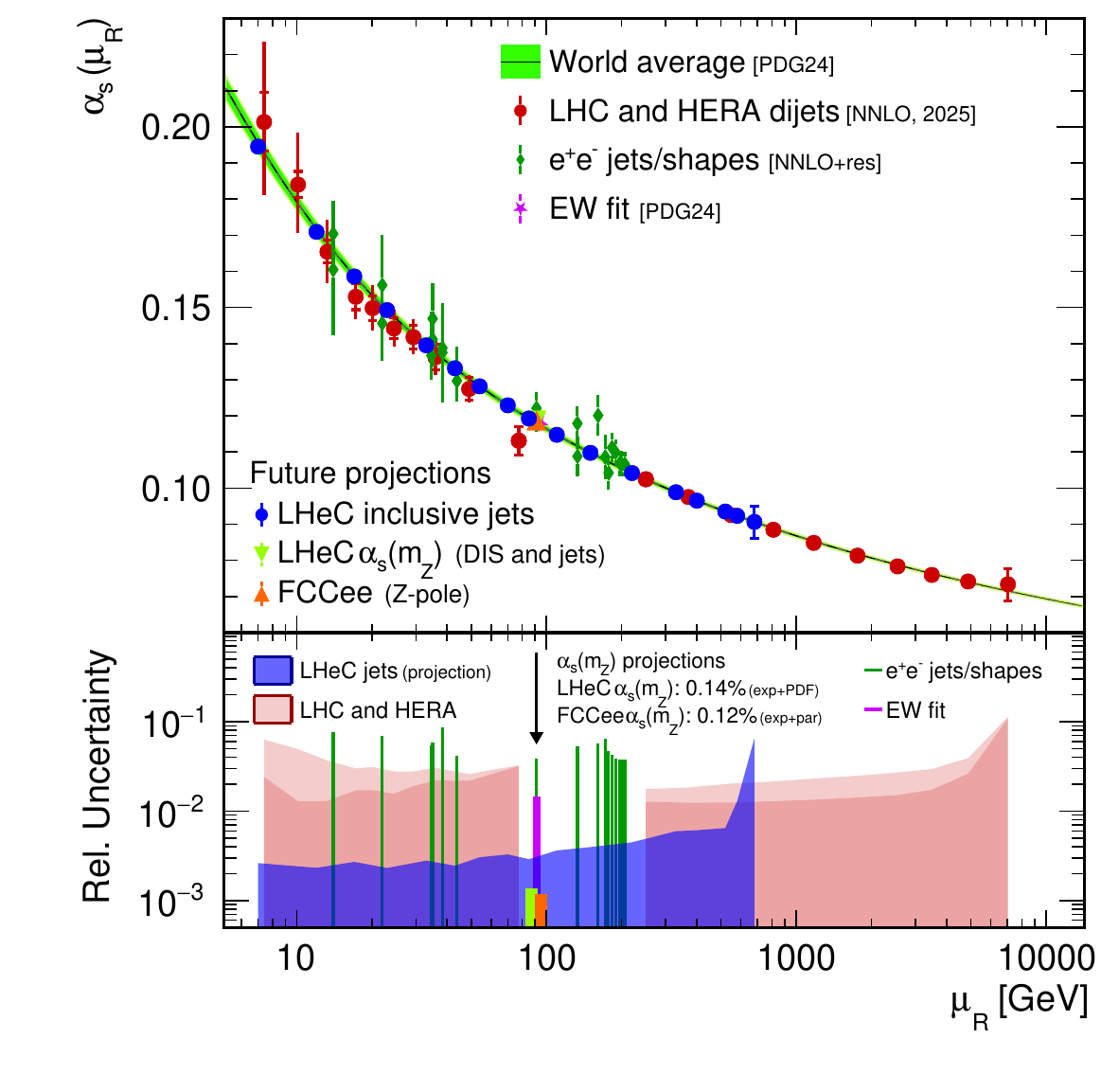}
\caption{Projected LHeC determinations of $\alpha_s(\mu_R)$ from inclusive jets and from the combined DIS-plus-jets determination, compared with present measurements and future projections. The lower panel shows the relative precision as a function of the renormalisation scale.}
\label{fig:alphas}
\end{figure}

Building on the HERA jet programme, including the NNLO H1 extraction of $\as(\mz)$~\cite{H1:2017bml}, the bridge-project study quotes a projected uncertainty from inclusive DIS of about $\delta\as(\mz)=0.00022$, and from inclusive jets of about $\delta\as(\mz)=0.00013$~\cite{Ahmadova:2025,Agostini:2020}. These uncertainties would improve substantially on present DIS jet determinations and would provide a stringent test of the running of the coupling over a wide scale range. Fig.~\ref{fig:alphas} shows the projected scale dependence together with present measurements and future expectations.

\section{Electroweak measurements}

Inclusive DIS at the LHeC is also a precision electroweak laboratory. At high $Q^2$, NC and CC cross sections receive sizeable $\gamma/Z$ interference, pure-$Z$ and $W$-exchange contributions, so simultaneous EW+PDF fits can determine light-quark neutral-current couplings, test charged-current normalisations, and extract the weak mixing angle in the spacelike regime~\cite{Britzger:2020kgg,Britzger:2022wuc,ParticleDataGroup:2024cfk}. This is complementary to $Z$-pole, fixed-target and hadron-collider determinations because it probes different initial states, momentum transfers and PDF correlations.

\begin{figure}[t]
\centering
\includegraphics[width=0.60\textwidth]{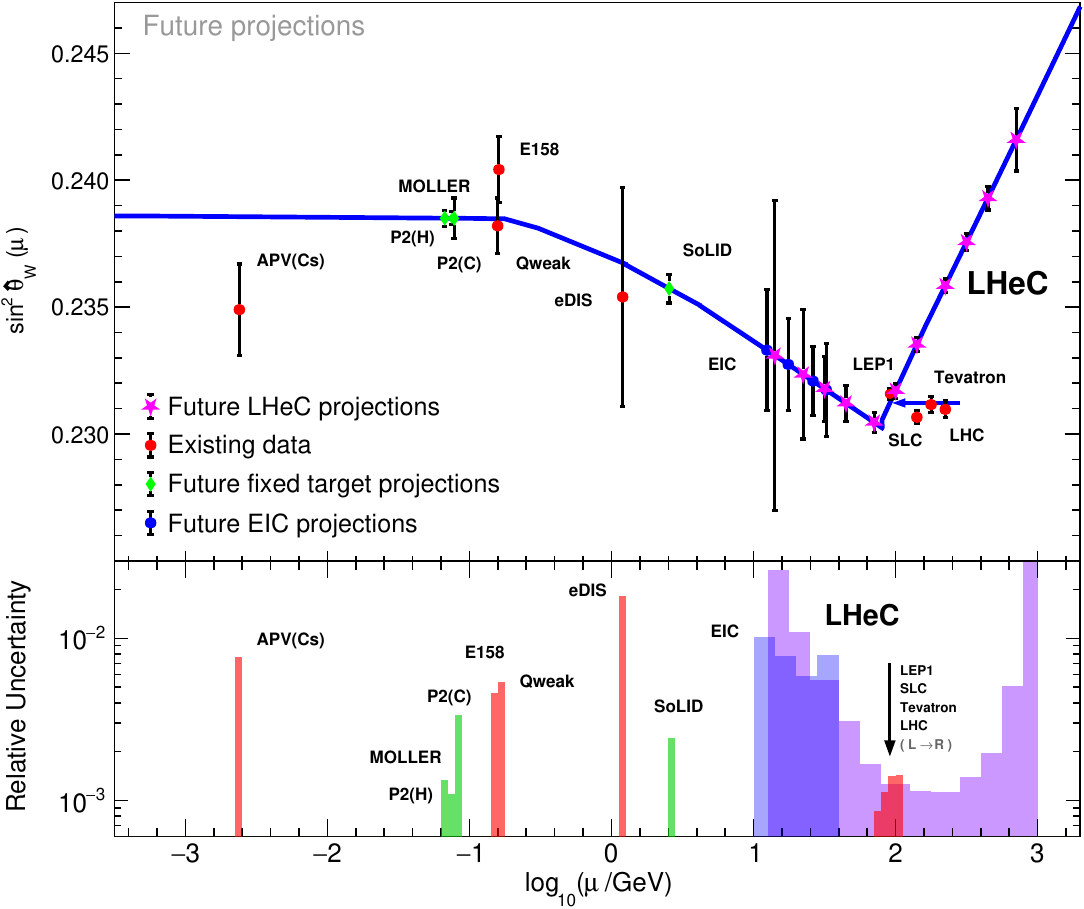}
\caption{Projected sensitivity to the scale dependence of the weak mixing angle. Existing data, future fixed-target and EIC projections are compared with LHeC points spanning the electroweak-scale region. The lower panel shows the corresponding relative uncertainties.}
\label{fig:sin2thetaW}
\end{figure}

Fig.~\ref{fig:sin2thetaW} shows the projected LHeC reach for $\sin^2\theta_W(\mu)$. The LHeC points cover the transition from intermediate DIS scales to the TeV domain and provide a direct test of the Standard Model running in one experimental environment. The precision is driven by the high luminosity, both electron charges and polarisations, and the PDF constraints discussed above; conversely, a controlled EW fit is required for unbiased high-$Q^2$ PDF extractions and BSM interpretations.

\section{Small-$x$ QCD, heavy flavours and high-density matter}

The low-$x$ reach in Fig.~\ref{fig:kinplane} is not only a PDF-improvement lever arm; it is also a discovery tool for QCD dynamics. In fixed-order collinear evolution the rapid growth of the gluon at small $x$ drives scaling violations of inclusive structure functions. At sufficiently high densities, gluon recombination and saturation effects may tame this growth. The LHeC can test this region through precise measurements of $F_2$, $F_L$, charm and beauty structure functions, jets and forward final states. Measurements of $F_L$ are particularly sensitive because they probe the gluon distribution and the longitudinal component of the exchanged boson.

\begin{figure}[t]
\centering
\begin{minipage}{0.48\textwidth}
\centering
\includegraphics[width=\linewidth]{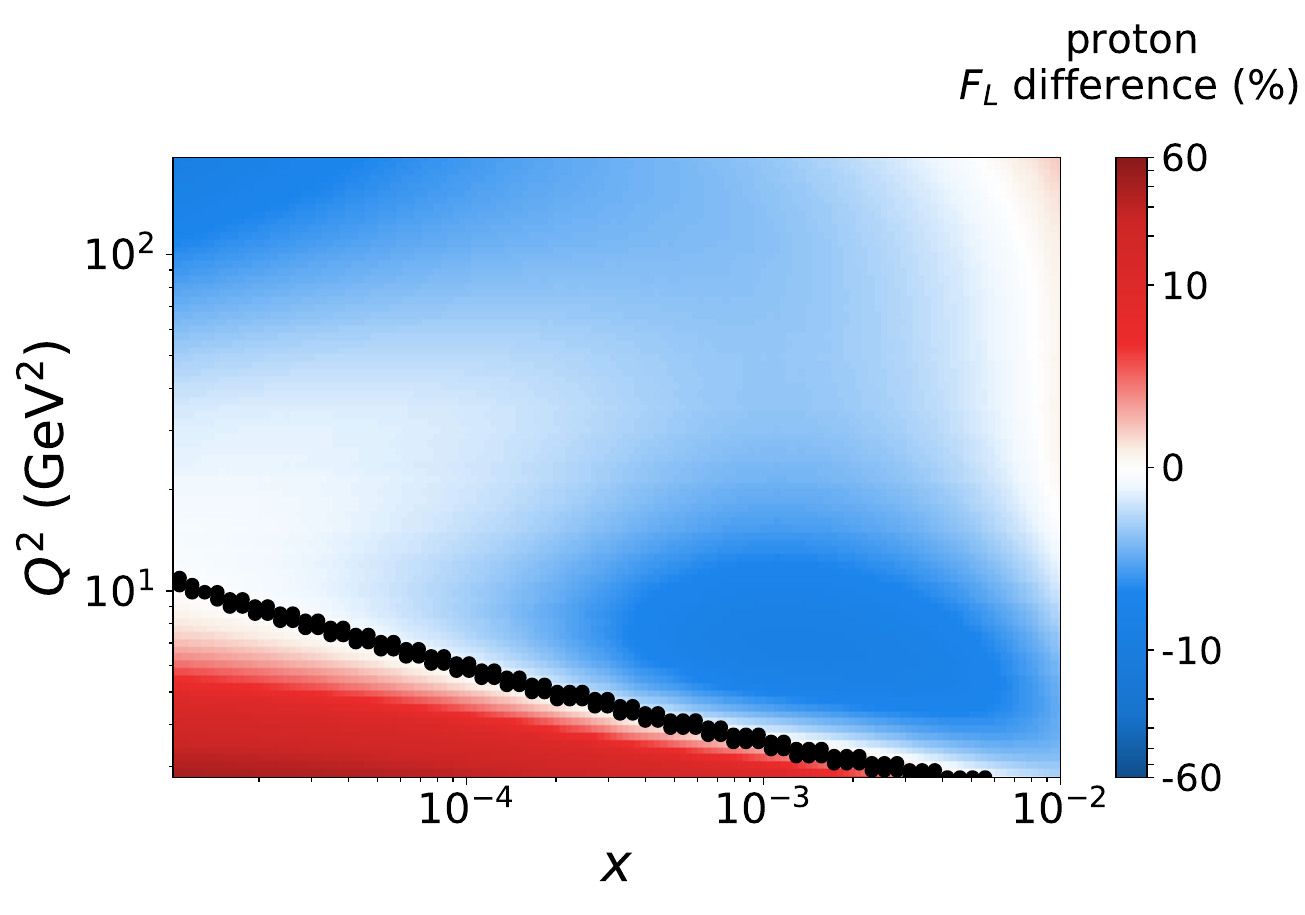}
\end{minipage}\hfill
\begin{minipage}{0.48\textwidth}
\centering
\includegraphics[width=\linewidth]{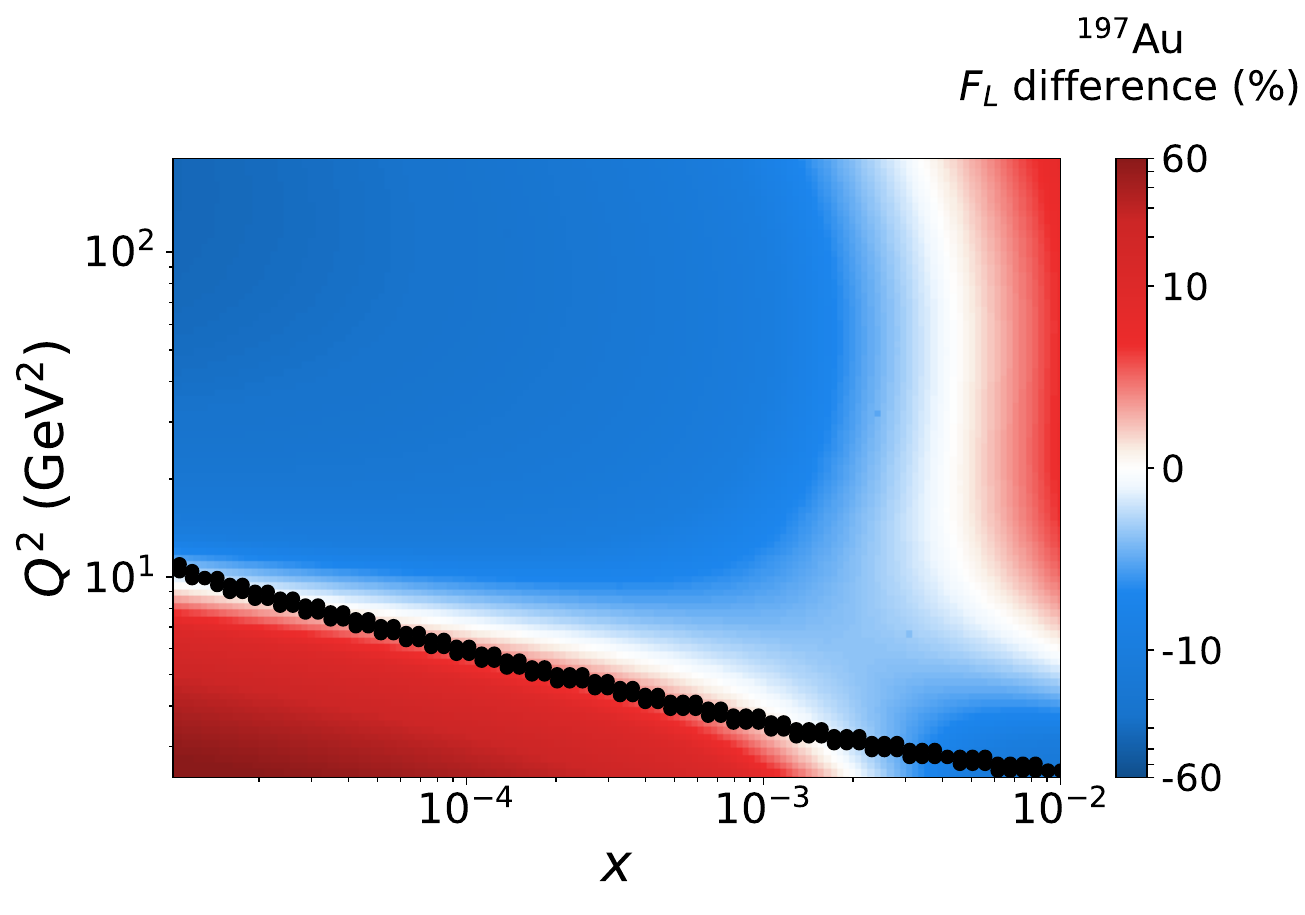}
\end{minipage}
\caption{Sensitivity of the longitudinal structure function $F_L$ in the high-energy, high-density region. The colour scale shows the relative difference in $F_L$ in the $(x,Q^2)$ plane for proton beams (left) and $^{197}{\rm Au}$ beams (right). The nuclear case enhances density effects and therefore provides a powerful lever arm for distinguishing linear small-$x$ evolution from non-linear QCD dynamics in electron--ion collisions.}
\label{fig:FLheavyions}
\end{figure}

An explicit extension of this programme is high-energy electron--ion scattering. With heavy nuclei the parton density is amplified, so the same low-$x$ kinematics probe a denser QCD environment than in $ep$. Fig.~\ref{fig:FLheavyions} compares projected $F_L$ sensitivity maps for the proton and for $^{197}{\rm Au}$, complementing the planned EIC nuclear-DIS programme~\cite{AbdulKhalek:2021gbh}. The changes are concentrated in the low-$x$, moderate-$Q^2$ region where non-linear effects are expected to be most visible, and the nuclear plot displays a broader and stronger pattern than the proton case. Such measurements would connect proton-PDF physics with the high-density initial conditions relevant for heavy-ion collisions, constraining nuclear PDFs, shadowing, and possible saturation effects with clean DIS observables.

The clean DIS environment also offers a complementary route to heavy-flavour PDFs. Charm and beauty production test heavy-quark mass schemes and perturbative matching conditions, while strange production can be constrained by semi-inclusive charm and light-hadron measurements. These measurements reduce PDF correlations that currently propagate into hadron-collider predictions, including $W$ and $Z$ production, Higgs production, and high-mass tails.

\section{Conclusions}

The LHeC would determine proton and nuclear PDFs, test high-density QCD with $F_L$ and heavy-flavour observables, measure $\as$, and provide spacelike electroweak precision tests in a single DIS programme. These results would reduce theory-systematic uncertainties for the HL-LHC and future hadron colliders, with significant impact already at $50\fb$ and transformative constraints at $1\ab$.

\begingroup
\scriptsize
\begin{multicols}{2}

\end{multicols}
\endgroup


\begin{thebibliography}{99}

\bibitem{Ahmadova:2025}
F.~Ahmadova et al.,
``The Large Hadron electron Collider as a bridge project for CERN,''
\href{https://arxiv.org/abs/2503.17727}{arXiv:2503.17727 [hep-ex]}.

\bibitem{Agostini:2020}
P.~Agostini et al. [LHeC and FCC-he Study Group],
``The Large Hadron--Electron Collider at the HL-LHC,''
J. Phys. G \textbf{48} (2021) 110501,
\href{https://arxiv.org/abs/2007.14491}{arXiv:2007.14491 [hep-ex]}.

\bibitem{LHeCStudyGroup:2012zhm}
J.~L.~Abelleira Fernandez et al. [LHeC Study Group],
``A Large Hadron Electron Collider at CERN,''
J. Phys. G \textbf{39} (2012) 075001,
\href{https://arxiv.org/abs/1206.2913}{arXiv:1206.2913 [physics.acc-ph]}.

\bibitem{Klein:2018rhq}
M.~Klein,
``Future Deep Inelastic Scattering with the LHeC,''
\href{https://arxiv.org/abs/1802.04317}{arXiv:1802.04317 [hep-ph]}.

\bibitem{PDF4LHCWorkingGroup:2022cjn}
R.~D.~Ball et al. [PDF4LHC Working Group],
``The PDF4LHC21 combination of global PDF fits for the LHC Run III,''
J. Phys. G \textbf{49} (2022) 080501,
\href{https://arxiv.org/abs/2203.05506}{arXiv:2203.05506 [hep-ph]}.

\bibitem{NNPDF:2021njg}
R.~D.~Ball et al. [NNPDF],
``The path to proton structure at 1\% accuracy,''
Eur. Phys. J. C \textbf{82} (2022) 428,
\href{https://arxiv.org/abs/2109.02653}{arXiv:2109.02653 [hep-ph]}.

\bibitem{H1:2015ubc}
H.~Abramowicz et al. [H1 and ZEUS],
``Combination of measurements of inclusive deep inelastic $e^{\pm}p$ scattering cross sections and QCD analysis of HERA data,''
Eur. Phys. J. C \textbf{75} (2015) 580,
\href{https://arxiv.org/abs/1506.06042}{arXiv:1506.06042 [hep-ex]}.

\bibitem{Hou:2019efy}
T.~J.~Hou et al.,
``New CTEQ global analysis of quantum chromodynamics with high-precision data from the LHC,''
Phys. Rev. D \textbf{103} (2021) 014013,
\href{https://arxiv.org/abs/1912.10053}{arXiv:1912.10053 [hep-ph]}.

\bibitem{Bailey:2020ooq}
S.~Bailey et al.,
``Parton distributions from LHC, HERA, Tevatron and fixed target data: MSHT20 PDFs,''
Eur. Phys. J. C \textbf{81} (2021) 341,
\href{https://arxiv.org/abs/2012.04684}{arXiv:2012.04684 [hep-ph]}.

\bibitem{Alekhin:2017kpj}
S.~Alekhin, J.~Bl\"umlein, S.~Moch and R.~Placakyte,
``Parton distribution functions, $\alpha_s$, and heavy-quark masses for LHC Run II,''
Phys. Rev. D \textbf{96} (2017) 014011,
\href{https://arxiv.org/abs/1701.05838}{arXiv:1701.05838 [hep-ph]}.

\bibitem{H1:2017bml}
V.~Andreev et al. [H1],
``Determination of the strong coupling constant $\alpha_s(m_Z)$ in next-to-next-to-leading order QCD using H1 jet cross section measurements,''
Eur. Phys. J. C \textbf{77} (2017) 791,
\href{https://arxiv.org/abs/1709.07251}{arXiv:1709.07251 [hep-ex]}.

\bibitem{Britzger:2020kgg}
D.~Britzger, M.~Klein and H.~Spiesberger,
``Electroweak Physics in Inclusive Deep Inelastic Scattering at the LHeC,''
\href{https://arxiv.org/abs/2007.11799}{arXiv:2007.11799 [hep-ph]}.

\bibitem{Britzger:2022wuc}
D.~Britzger, M.~Klein and H.~Spiesberger,
``Precision electroweak physics at the LHeC and FCC-eh,''
\href{https://arxiv.org/abs/2203.06237}{arXiv:2203.06237 [hep-ph]}.

\bibitem{ParticleDataGroup:2024cfk}
S.~Navas et al. [Particle Data Group],
``Review of Particle Physics,''
Phys. Rev. D \textbf{110} (2024) 030001.

\bibitem{AbdulKhalek:2021gbh}
R.~Abdul Khalek et al.,
``Science Requirements and Detector Concepts for the Electron-Ion Collider: EIC Yellow Report,''
Nucl. Phys. A \textbf{1026} (2022) 122447,
\href{https://arxiv.org/abs/2103.05419}{arXiv:2103.05419 [physics.ins-det]}.

\end{thebibliography}
\end{document}